\documentclass[twocolumn,showpacs,preprintnumbers,amsmath,amssymb]{revtex4}

\usepackage{graphicx,epsf}
\usepackage{dcolumn}
\usepackage{bm}

\begin{document}

\title{Magneto-Stark polaron states in semiconductor quantum wells}

\author{Yu  Chen$^{1,2}$, Nicolas Regnault$^1$, Robson Ferreira$^1$}
\affiliation{$^1$Laboratoire Pierre Aigrain, Ecole Normale Sup\'erieure, CNRS (UMR 8551), Univ. P. et M.
Curie, Univ. D. Diderot ; 24, rue Lhomond, 75005, Paris, France\\
$ ^2$Department of Physics, Tsinghua University, Beijing 100084,
China}

\date{\today}

\begin{abstract}
We study theoretically the effect of a lateral electric field on the magneto-polaron
states in a quantum cascade laser under a quantizing magnetic field. We show that this problem fits the original
Fano model of a discrete state coupled to a continuum, present a detailed analysis of
magneto-Stark polaron resonances, and finally discuss the strong consequences on the optical characteristics
of an operating structure.
\end{abstract}


\maketitle

Nanostructures made of semiconductors are model objects for the study of
fundamental interactions in solid-state physics. This results from both the large
quantization of their electronic levels and their peculiar response to external
fields. Moreover, progresses in growth techniques brought about the concept of
spectral engineering, where by a proper multilayer design one is able to
construct a desired sequence of levels. Quantum cascade lasers (QCL), which
perfectly illustrate this concept, have attracted intense research since their
discovery: in there, one realizes a sequence of states allowing population
inversion in a region made of a few layers of wells and barriers (the active
region) ~\cite{Faist}. More recently, it has been shown that a strong magnetic field
applied parallel to the growth axis of a QCL structure modifies its emission
characteristics ~\cite{Ulich}. This effect has been associated to the existence of field-induced
resonances between Landau levels (LL) of different subbands involved
in the population inversion. Inelastic resonances have also been studied, where
the magnetic field is chosen so that the two LL's are not resonant but detuned by
the energy of one longitudinal optical (LO) phonon ~\cite{Becker02}. For such field
values, one predicts the formation of resonant magneto-polarons (MP) states
~\cite{Becker04, Chen}, which are entangled eigenstates issue from the strong coupling between
electrons and LO phonons (by the Fr\"olich interaction, prevalent in binary
materials often used to fabricate QCL's, like GaAs and InAs). In this work we
account for an additional electric field applied perpendicularly to the magnetic
field. This is the well-known fields configuration for Hall (classical or
quantum) effect ~\cite{2DEG}. However, we are not interested here in the low-lying
states, but shall discuss the effect of the external bias on the highly excited
MP states involving the upper lasing level. As we show, this is a model
problem, with interest both on the fundamental and applicative sides. Indeed,
we demonstrate that it fits the Fano model of a discrete state coupled to a
continuum ~\cite{Fano}. We demonstrate that the interaction between electrons and LO
phonons can be made to evolve continuously from a strong to a weak coupling
regime by increasing the electrostatic field. We finally discuss the consequence
of this striking qualitative and quantitative bias-triggered evolution of the energy
eigenstates on the optical absorption of a QCL.

Our aim is to describe the magneto-Stark polaron (MSP) states that follow from
the interplay of the crossed fields and electron-phonon coupling, and their
influence on the QCL optical transitions. For a qualitative rather than a
quantitative understanding of the concomitant effects of these three couplings on
the QCL functioning, we model its active region by a single quantum well with
infinitely height barriers. To start with, let us briefly recall the cases of magneto-
Stark states in crossed $F$ and $B$ (in absence of electron-phonon coupling). We
consider a quantizing magnetic field applied along the growth $Oz$ direction,
described in the Landau gauge ${\vec A=Bx\hat{y}}$, and an electric field applied along the
$Ox$ direction. This problem is exactly solvable, with eigenstates~\cite{Helm}:
\begin{eqnarray}
\big<\vec
r\big|E_l,n,k_y\big>&=&\frac{1}{\sqrt{L_y}}e^{ik_yy}\chi_l(z)\varphi_n(x+\lambda^2k_y-\frac
{v_{D}}{\omega_{c}}),{}
\nonumber\\
{}\varepsilon_{l,n,k_{y}}&=&E_{l}+\Big(n+\frac{1}{2}\Big)\hbar\omega_{c}-eF\lambda^2
k_{y}-\frac{1}{2}m^{*}v_{D}^{2}
\nonumber
{}
\end{eqnarray}
where $m^{*}$ is the electron effective mass,
${\lambda=\sqrt{\hbar/{eB}}}$ the magnetic length, ${v_{D}=F/B}$ the drift velocity,
$\omega_{c}=eB/m^{*}$ the cyclotron frequency, $\chi_l$ the $l$-th QW
subband wavefunction, $k_y$ the wavevector along the $Oy$ direction (of
macroscopic size $L_y$) and ${\varphi_n}$ the $n$-th Hermite function. We consider a QW
width $L_z=12.33nm$, which gives a laser emission around 8.7 nm between the
first two subbands of a GaInAs-based structure ($m^{*}=0.05m_0$ and $E_{2}-E_{1}=147.5meV$).
As it is well known, the electric field lifts the LL
degeneracy related to the plane-wave motion along $Oy$, generating an ensemble
of crossed magneto-Stark levels, and induces a drift motion of the electron
perpendicular to both fields.

\begin{figure}
\includegraphics[height=.25\textheight, angle=270]{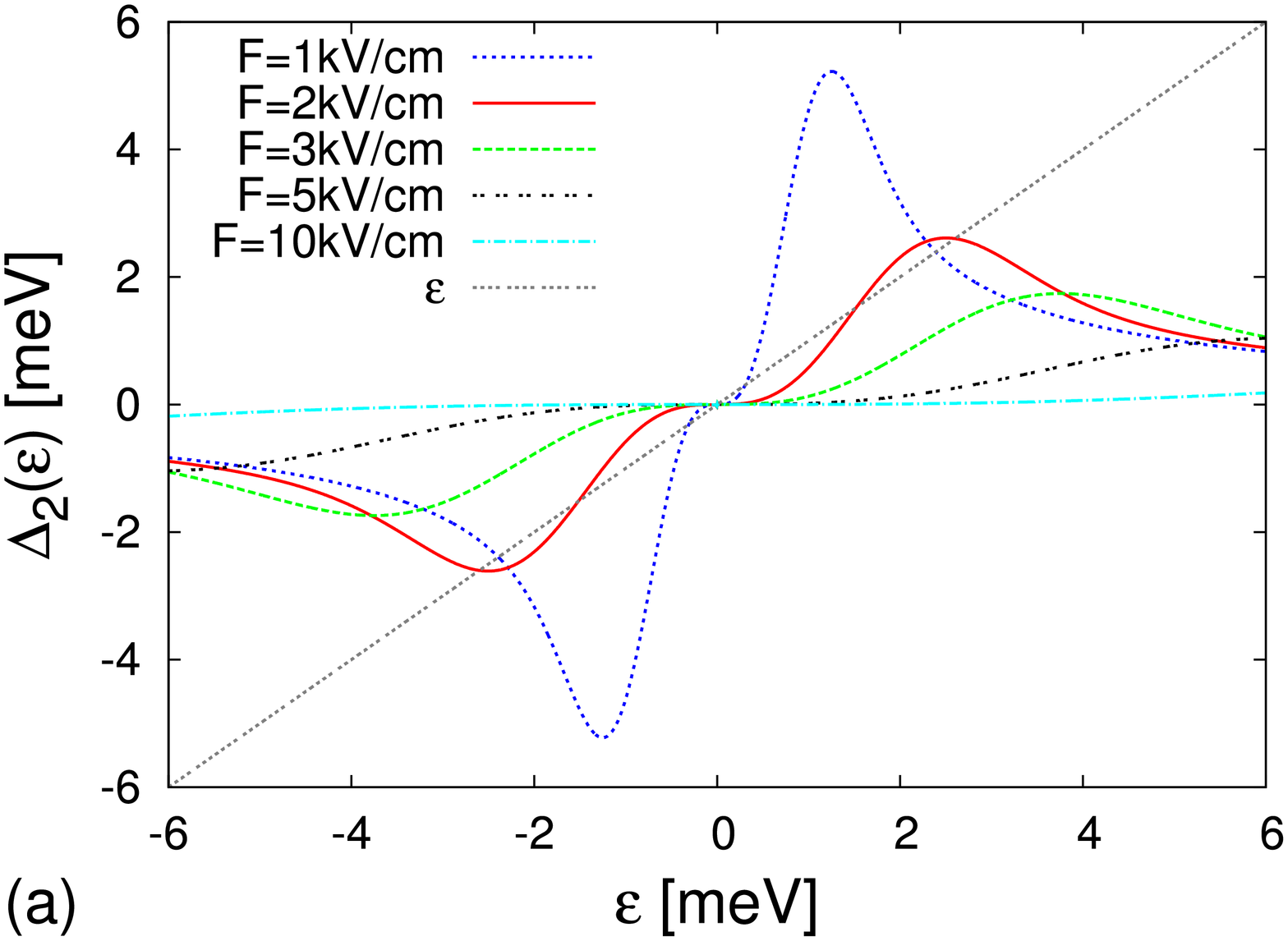}\\
\includegraphics[height=.25\textheight, angle=270]{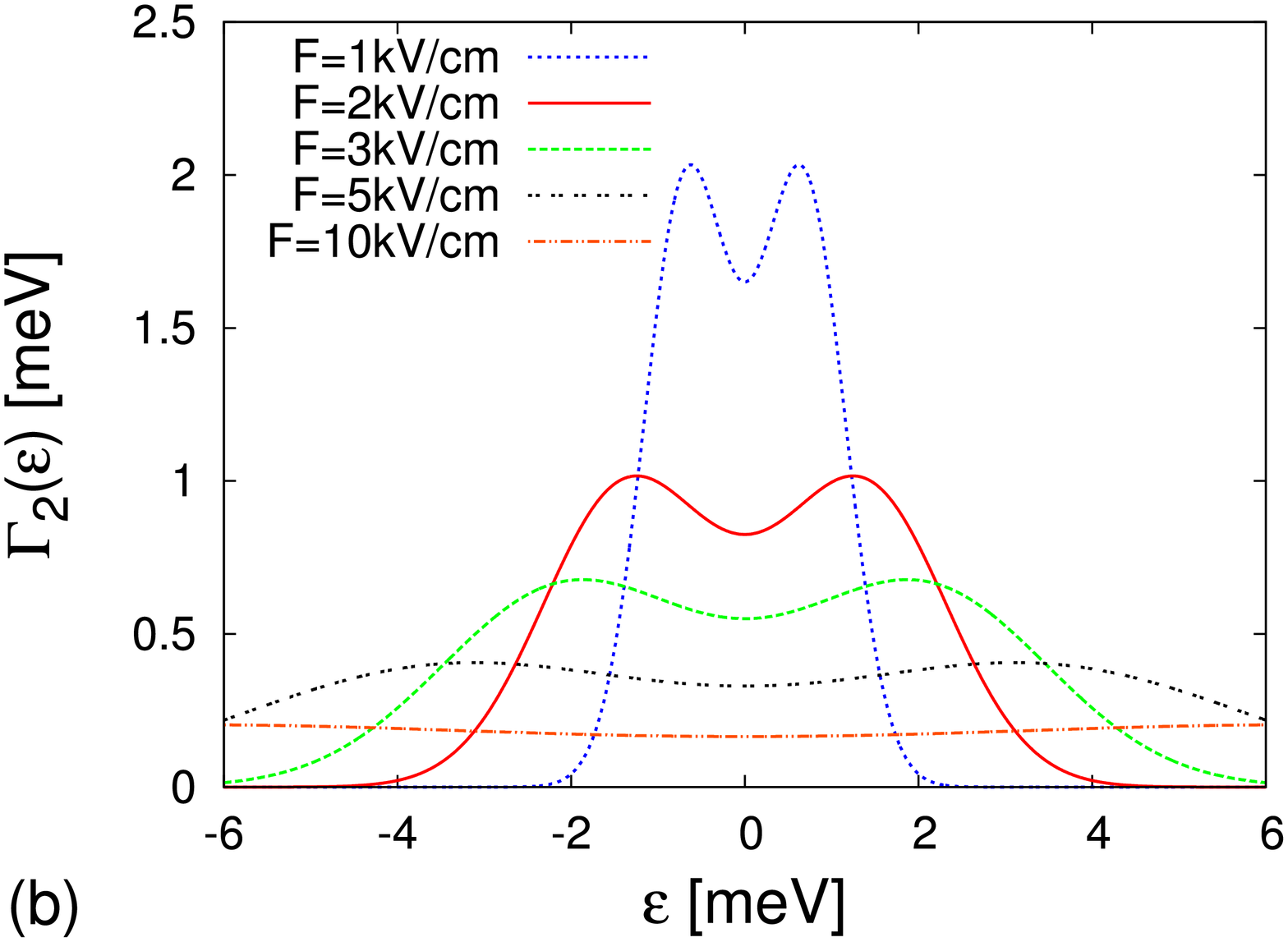}
  \caption{(Color online) Functions ${\Delta_2}$ (a) and ${\Gamma_2}$ (b) as functions of the energy
  ${\varepsilon=E-E_d}$ for different values of the electric field $F$, at the $p=2$ resonance (${B_2=24.6T}$).}
\label{delta2fig}
\end{figure}

\begin{figure}
\includegraphics[height=.25\textheight, angle=270]{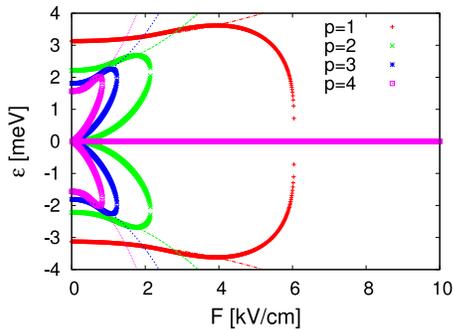}
\caption{(Color online) energy position of the Magneto-Stark polaron resonances as
functions of the external electric field, for a few magnetic fields ${B_p}$ (${p=1}$-${4}$;
${\varepsilon=E-E_{d}}$). Broken lines: low electric field approximation (see text).}
\label{energypos}
\end{figure}

In this work we are interested in the effect of the electrostatic field on the magneto-polaron (MP) 
states. Indeed, it has been shown in previous works, done at
$F=0$ ~\cite{Becker04, Chen}, that electrons in a QW subjected to a high magnetic field strongly couple
to Longitudinal Optical (LO) phonons by the Fr\"ohlich interaction~\cite{Kittel}
${{V_{e-ph}=\sum_{\vec q}\Big(V_{q}e^{-i\vec q\cdot\mbox{}\vec
r}b_{\vec q}^{+}+h.c.\Big)}}$ , where ${b_{\vec q}^{+}}$
is the creation operator for LO phonons
with wavevector ${\vec q}$ and ${V_{q}=i\sqrt{{e^2\hbar\omega_{LO}}(\varepsilon_{\infty}^{-1}-\varepsilon_{s}^{-1})/{(2V_{cr}\varepsilon_{0}q^2})}}$
with $V_{cr}$ the crystal volume. We assume dispersionless phonons with energy $\hbar\omega_{LO}=33.7meV$, and
take ${\varepsilon_{\infty}=11.1}$ and ${\varepsilon_{s}=13.1}$ for the high-frequency and static relative dielectric
constants, respectively. At $F=0$, MP states are resonantly
formed at the fields $B_{p}=m^{*}(E_2-E_1-\hbar\omega_{LO})/(pe\hbar)=B_1/p$, where the $p$-th
($p=1,2,...$) LL from the $E_1$ subband with one LO phonon occupancy crosses the
$n=0$ LL of the $E_2$ one. Following the development of ~\cite{Becker04, Chen}, we look for MP
states around $B=B_p$ at $F\neq0$ in the form:
\begin{eqnarray}
\big|\Psi\big>&=&a_d\big|\varphi_d\big>+\sum_{q_y}b_{q_y}\big|\nu_{q_y}\big>,
\end{eqnarray}
where 
\begin{eqnarray}
\big|\varphi_d\big>=\big|E_2,0,k_y\big>\otimes\big|0_{LO}\big>,&& E\left(\big|\varphi_{d}\big>\right)\equiv E_{d}=\varepsilon_{2,0,k_y}
\nonumber
\end{eqnarray}
\begin{eqnarray}
\big|\nu_{q_y}\big>&=&\frac{\sum_{q_x,q_z}{\alpha_{q_x,q_z}(k_y,q_y)\big|E_1,p,k_y-q_y\big>\otimes\big|1_{\vec q}\big>}}{\sum_{q_x,q_z}{\left|\alpha_{q_x,q_z}(k_y,q_y)\right|^2}},
\nonumber
\end{eqnarray}
\begin{eqnarray}
\alpha_{q_x,q_z}(k_y,q_y)&=&V_{q}^{*}\big<E_{1},p,k_y-q_y\big|e^{-\vec q\cdot\vec r}\big|E_2,0,k_y\big>
\nonumber\\
&=&V_q^{*}\big<\chi_{1}(z)\big|e^{-iq_{z}z}\big|\chi_{2}(z)\big>
\nonumber\\
&&\times\big<\varphi_{p}(x-\lambda^{2}q_{y})\big|e^{-iq_{x}x}\big|\varphi_{0}(x)\big>\label{defalpha}\\
&&\times e^{iq_{x}(\lambda^{2}k_{y}-v_{D}/\omega_{c})},
\nonumber\\
E\left(\big|\nu_{q_y}\big>\right)&\equiv& E_{q_y}=\varepsilon_{1,p,k_y-q_y}+\hbar\omega_{LO}
\nonumber\\
&=&E_{d}-\Delta E+eF\lambda^{2}q_{y},
\nonumber
\end{eqnarray}
whit $\Delta E=E_{2}-(E_{1}+p\hbar\omega_{c}+\hbar\omega_{LO})$ is the $B$-dependent detuning ($\Delta E(B_p)=0$).
We have used explicitly the fact that $k_y$ remains a good quantum number for the
magneto-Stark polaron (MSP) states, and dropped it from the labelling of $\big|\Psi\big>$
and the coefficients $a$ and $b$. Thus, for a fixed $k_y$ value, we have a Fano problem
of coupling (by the Fr\"ohlich interaction) of a discrete state (of energy $E_d$) to a
field-induced continuum (centered at $E_d-\Delta E$ and of spectral width $W=2\pi\hbar v_{D}/a_{0}$,
since $q_y$ varies continuously in the interval $(-\pi/a_0,\pi/a_0)$, with $a_0$ the crystal unit
cell) . This problem exactly fits the one treated by U. Fano in its original work
~\cite{Fano}. The energies fulfil
\begin{eqnarray}
E=E_{d}+\Delta_{p}(E)+z_{p}(E)\Gamma_{p}(E),
\nonumber\\
\Delta_{p}(E)={\rm P}\int dq_{y}\frac{K_{p}(q_y)}{E-E_{q_y}},
\nonumber\\
\Gamma_{p}(E)=\int dq_{y} K_{p}(q_y)\delta\left[E-E_{q_y}\right],
\nonumber\\
K_{p}(q_y)=\frac{L_{y}}{2\pi}\sum_{q_x,q_z}\left|\alpha_{q_x,q_z}(k_y,q_y)\right|^2,
\end{eqnarray}
where $\rm P$ is the principal part and the phase-shift term $z_{p}(E)$ is defined from the
first equation. We show in figure 1 the functions $\Delta_{p}(E)$ (figure \ref{delta2fig}(a)) and $\Gamma_{p}(E)$
(figure \ref{delta2fig}(b)) for $B_{2}=24.6T$ and different electric fields. This figure shows that, 
even though the spectral width of the $F$-induced
continuum can be very large as compared to the cyclotron energy ($W\approx
10\hbar\omega_c$ for $F=2 kV/cm$ and $B_2$), its effective spectral width, as dictated by the
energy range where the matrix elements $K_p(q_y)$ is sizeable, is much narrower for
low fields. This justifies using in the construction of $\big|\Psi\big>$ only states related to
the two quasi-resonant LL's. The energy positions of a Fano resonance fulfills $E-E_d=\Delta_{p}(E)$.
As can be easily seen graphically in figure \ref{delta2fig}(a), there are actually five
solutions at low fields (up to $\approx 2kV/cm$) : one at $E=E_d$ and four resonances
symmetrically disposed around $E_d$. We plot in figure \ref{energypos} the
energies of the MSP resonances as a function of the electric field, for various
magnetic fields $B_p$. One can show that $K_p(q_y)$ is an even function of $q_y$; it
straightforwardly results that there is an odd number of resonances : $E=E_d$ is
always a solution for $F\neq0$, and there is possibly also an even number of
solutions, disposed symmetrically with respect to the energy origin. The
solution at the origin is the only possible at high fields, i. e., above a critical
field $F_p$ that increases with increasing $B_p$ . Below this
critical field, we obtain additionally two (for $p=1$) or four (for $p \geqslant 2$) non-trivial
MSP resonances, which display a ``butterfly"-like shape when plotted as a
function of $F$. In the following we discuss the origin of this butterfly spectrum
and show that it follows from two main ingredients: (i) the passage from a
strong to a weak electron-phonon coupling regime with increasing electric field,
and (ii) the energy-dependence of the matrix elements $K_p(q_y)$.

In order to understand the previous results, it is worth recalling that for
$F=0$ the continuum is flat and the electron-phonon interaction is by essence
strong. As shown in ~\cite{Becker04, Chen}, at resonance field $B_p$, its diagonalization generates a
highly degenerated set of non-interacting one-phonon states at the energy $E_d$ and
two split MP levels at energies $E_d\pm \hbar\Omega_p$, where $\hbar\Omega_p$ is the polaron coupling
strength. This latter represents the average coupling to the flat continuum and
increases with magnetic field: ${\Omega_p\propto\sqrt{B_p}}$. However, some states of the
continuum enter with more weight in the formation of the MP state, as becomes
clear by considering the $x$-related integral in Eq. (\ref{defalpha}) (let us take $q_x=0$, for
simplicity): the overlap of the two displaced harmonic oscillators wavefunctions
is maximum when the displacement is of the order of the extension of their
wavefunctions, i.e., $\lambda^{2}q_y\approx\sqrt{2p+1}\lambda$. The external bias gives a finite width to the
continuum by shifting differently states with different $q_y$ values (by
$E_{q_y}-E_d=eF\lambda^{2}q_y$). One expects then that the strong coupling regime at the origin of the
$F=0$ magneto-polaron states survives to the applied bias provided that the states
with larger coupling strengths are not significantly shifted, i.e., for
$eF_p\lambda^{2}q_y\leqslant\hbar\Omega_p$. These two results lead to
$eF_p\lambda\sqrt{2p+1}\approx\hbar\Omega_p$, which has a
simple physical interpretation: the critical electric field is such that the
electrostatic potential drop along the Landau orbit equals the average electron-phonon
interaction. We have checked that this simple analysis is in very good agreement with the full
numerical calculus of $F_p$. Finally, at low
fields the energy positions of the MP resonances are given by:
$\hbar\Omega_p\left[1+(2p+1)\big(eF\lambda/(2\hbar\Omega_p)\big)^{2}\right]$;
these quadratic shifts are also plotted in figure \ref{energypos}.

The previous analysis explains the existence of a critical electrical field
and the initial quadratic shift of the MP state in figure \ref{energypos}, but does not explain the
existence of a vanishing waist in the butterfly profile for $p \geqslant 2$. The lower branch
follows from the particular form of the matrix elements $K_p(q_y)$. Indeed, around $B_p$, one
has approximately (using $Q=\lambda q_y$):
\begin{eqnarray}
K_{p}(Q)=\lambda\frac{\hbar^2\Omega_{p}^{2}}{2^{p+1}p!\sqrt{2\pi}}
{\left(\sum_{n=0}^{p-1}{G(p,n)Q^{2n}}+Q^{2p}\right)e^{-Q^{2}/2}},
\nonumber\\
G(p,n)=\frac{p!(2(p-n-1))!(p-n)(4p-4n-1)}{2^{p-n-1}n!((p-n)!)^2}.
\nonumber
\end{eqnarray}
Thus, $K_p$ is a continuously decreasing function of $Q$ for $p=1$, but presents a
camel-back shape for $p \geqslant 2$ (see the $\Gamma_2(E)$ curves in figure \ref{delta2fig}(b). 
This gives rise to a linear (for $p=1$) and to a slower (sub-linear for $p \geqslant 2$)
increase of $\Delta_p(E)$ near the energy origin, and thus to the additional butterfly
branches that extrapolate to $E_d$ when $F \rightarrow 0$. These additional resonances thus
reflect the internal structure of the coupling strength, as opposed to a
structureless matrix element often used in problems involving a discrete state
coupled to (and placed inside) a continuum.

Let us now consider the high field ($F>F_p$) situation. In this case, the
electron-phonon interaction can be treated in the weak coupling formalism. The
discrete state acquires a finite lifetime $\tau_p$, given as:
\begin{equation}
\frac{1}{\tau_{p}(F)}=\frac{2\pi}{\hbar}\Gamma_{p}(E_d)=\frac{\sqrt{2\pi}}{eF\lambda}\frac{\hbar^2\Omega_{p}^2}{2^{p+1}p!}G(p,0),\nonumber
\end{equation}
a result that can of course also be obtained by taking $V_{e-ph}$ as a perturbation in
the Fermi's golden rule. It represents the dissociation, triggered by the electron-phonon
coupling, of the zero-phonon initial state into the continuum of one-phonon
states spanned by the electrostatic field. In conclusion, although not
acting directly on the phonon degrees of freedom, the external bias affects the
strong electron-phonon coupling at the origin of the MP states in QCL's at high
magnetic fields, by rendering the latter resonances at weak fields, and allowing
the irreversible emission of one phonon at high fields. This continuous passage
from the strong to the weak coupling regimes is thoroughly handled by our non-perturbative
model.

In order to demonstrate the influence of these results on the optical
properties of a QCL structure, let us consider the absorption of light by electrons
in the lowest LL of the ground subband, i.e., in the states
$\big|E_1,0,k_y\big>$. For light of frequency $\nu$ propagating in the plane layer
and polarized along the $Oz$ axis (the usual configuration in operating QCL's),
the absorption coefficient towards the high-energy MSP states $\big|\Psi\big>$ is proportional to
$\left|a_d[E=E_d+h(\nu-\nu_{QW})]\right|^2$, where
\begin{equation}
|a_d(E)|^2=\frac{\Gamma_{p}(E)}{[E-E_d-\Delta_{p}(E)]^2+[\pi\Gamma_{p}(E)]^2},
\nonumber
\end{equation}
and $\hbar\nu_{QW}=E_2-E_1$ (this is because of the selection rules of the dipolar coupling,
which preserves both the electron in-plane degrees of freedom and the phonon
states ~\cite{GB}). In absence of electron-phonon coupling and the absorption
profile is a delta like centered at $\nu=\nu_{QW}$. We show in figure 3 the evolution of
the absorption intensity as a function of the light detuning $\nu-\nu_{QW}$ and the
electric field, at the $B_2$ resonance. The electron-phonon interaction and the
applied bias enormously affect the absorption profile. We clearly see at very
low fields two marked absorption peaks related to the MP states at $F=0$, which
increasingly broaden with increasing $F$ and becomes a faint trace for $F>F_p$, and
a single line around the energy origin, which is very broad and weak for $F<F_p$
but gain intensity and sharpens considerably with increasing $F$.

\begin{figure}
\includegraphics[height=.25\textheight, angle=270]{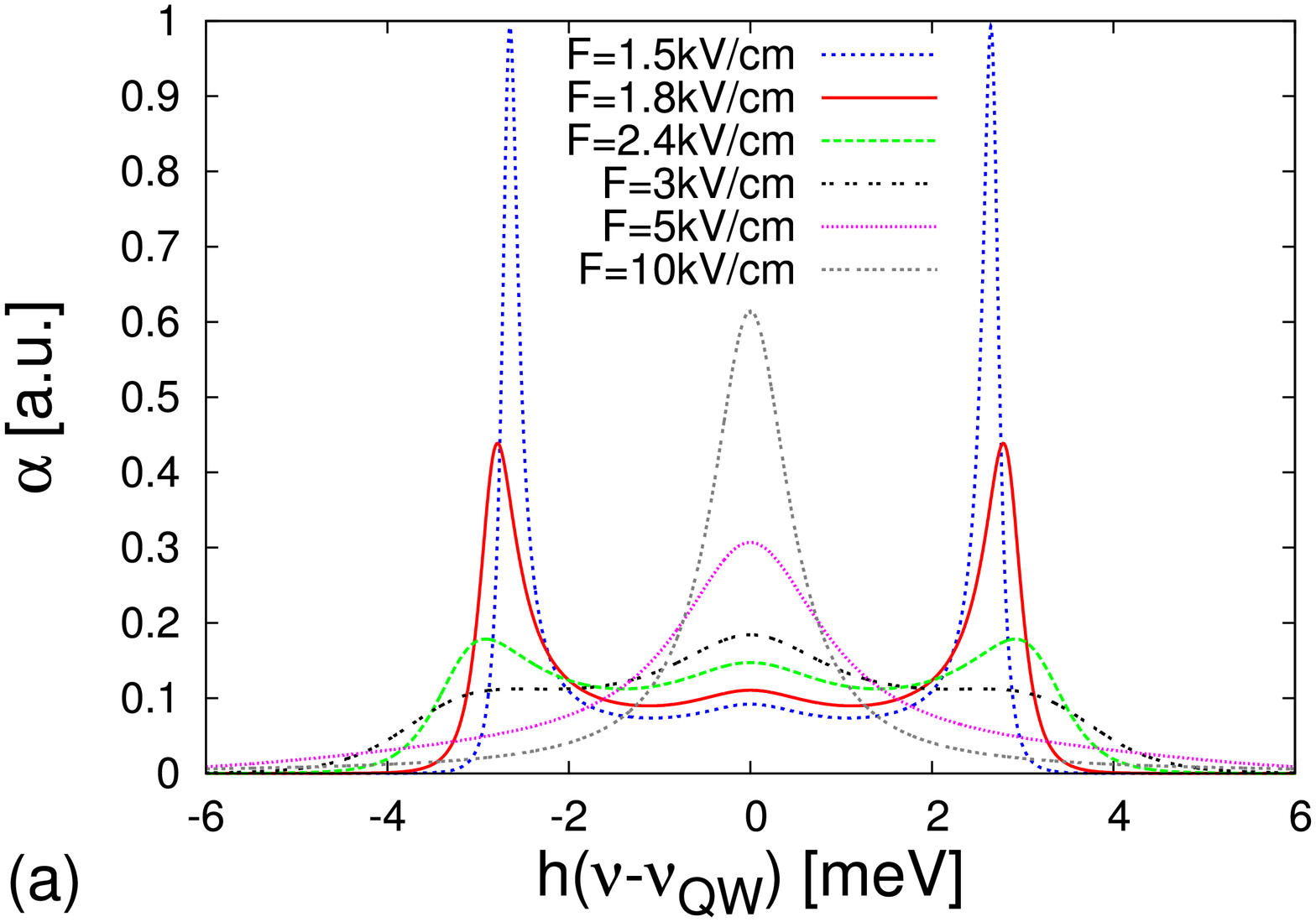}
\includegraphics[height=.25\textheight, angle=270]{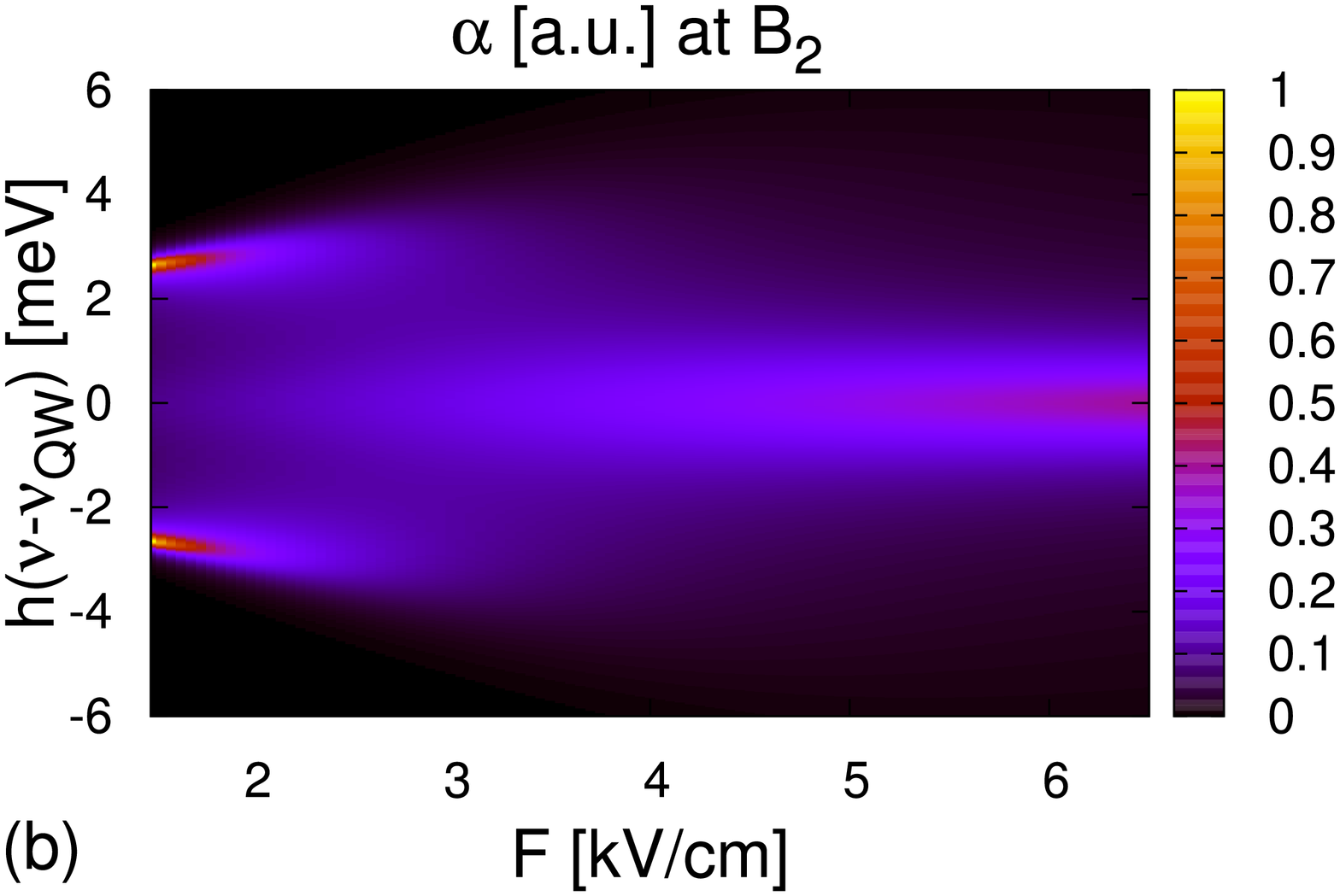}
  \caption{(Color online) Optical absorption spectra at ${B_2=24.6T}$. (a): energy spectra for different
  in-plane electric fields. (b): intensity-plot evolution with the applied bias. }
\end{figure}

Let us finally quote some additional remarks. First, the calculations were
performed for an InGaAs-based structure, which suffers from an important alloy
broadening~\cite{Chen}. However, the same phenomena are expected for GaAs-based
samples, with similar material parameters but sensibly free of alloy disorder.
Second, it is worth pointing out that the polaron coupling, and thus the $F$-induced
shift in the absorption spectrum, is in the few $meV$ range for the low $p$
resonances, and thus in principle large enough to be observed in actual samples.
Third, the critical field is in $kV/cm$ range, which is high enough to ensure the
stability of MP's against unavoidable local micro-field fluctuations in actual
samples, and at the same time low enough to prevent important lateral drift
during the dwell time of carriers in the active region of an operating QCL. Fourth,
we have also calculated the MSP states generated near (but not exactly at) the
resonance fields $B_p$, and obtained that the absorption line shape is affected in a
sizeable $B$-interval around $B_p$ (not shown). Finally, our results suggest that one
can sensibly monitor the rate of energy relaxation in the core region of a QCL,
by the application of an external modest lateral bias : indeed, MP states are
stationary entities that relax only by high-order (anharmonic~\cite{Grange}) processes,
comparatively much slower than a direct emission of one LO phonon (few $ps$
versus sub-$ps$ time scales, respectively~\cite{Grange}).

In conclusion, we have considered the magneto-Stark polaron states that
result from the interplay of crossed magnetic and electric fields and the electron-phonon
coupling in a semiconductor QW. This problem fits exactly the original
Fano problem of a discrete state coupled to a continuum, and thus admits an
exact solution, in a truncated basis spanned by crossing zero and one-phonon
states associated to LL's pertaining to different electron subbands. The model is
non-perturbative in either electrostatic and Fr\"olich interactions, and describes
in an unified way the continuous evolution from a strong to a weak coupling
regimes with increasing bias, i.e., the formation of MSP states, which are Fano-like
resonances at low fields, and their disintegration at higher fields. Finally,
we considered one consequence of such intricate interplay of different couplings
on the optical response (absorption spectrum) of a QCL structure and critically
considered its possible observation in actual samples.

We sincerely acknowledge Dr. G. Bastard and Prof. B.-F.
Zhu for many stimulating discussions. Y. C. thanks the French Ministry of
Foreign Affairs, the NSFC (Grant No.10774086) and the Basic Research
Program of China (grant 2006CB921500) for financial support.

\end{document}